\documentclass[aps,pra,twocolumn,floatfix,superscriptaddress]{revtex4-1}
\usepackage{newtxtext,newtxmath}
\usepackage{mathrsfs}
\usepackage{dsfont}
\usepackage{siunitx}
\usepackage[export]{adjustbox}
\usepackage[usenames,dvipsnames]{color}
\usepackage[colorlinks=true,citecolor=blue,breaklinks=true,linkcolor=red,
            linktocpage=true,urlcolor=blue,pagebackref=false]{hyperref}
\makeatletter
\newsavebox{\@brx}
\newcommand{\llangle}[1][]{\savebox{\@brx}{\(\m@th{#1\langle}\)}%
	\mathopen{\copy\@brx\mkern2mu\kern-0.9\wd\@brx\usebox{\@brx}}}
\newcommand{\rrangle}[1][]{\savebox{\@brx}{\(\m@th{#1\rangle}\)}%
	\mathclose{\copy\@brx\mkern2mu\kern-0.9\wd\@brx\usebox{\@brx}}}
\makeatother

\newcommand{\Tr}{\text{Tr}}

\begin{document}
\title{Dynamical multistability in a quantum dot laser}
\author{Mattia Mantovani}
\affiliation{Fachbereich Physik, Universit{\"a}t Konstanz,
             D-78457 Konstanz, Germany}
\author{Andrew D. Armour}
\affiliation{Centre for the Mathematics and Theoretical Physics of Quantum
             Non-Equilibrium Systems and School of Physics and Astronomy,
             University of Nottingham, Nottingham NG7 2RD, United Kingdom}
\author{Wolfgang Belzig}
\affiliation{Fachbereich Physik, Universit{\"a}t Konstanz,  D-78457 Konstanz,
             Germany}
\author{Gianluca Rastelli}
\affiliation{Fachbereich Physik, Universit{\"a}t Konstanz,  D-78457 Konstanz,
             Germany}
\affiliation{Zukunftskolleg, Universit{\"a}t Konstanz, D-78457 Konstanz,
             Germany}
\date{\today}
%
%
%
%
%
\begin{abstract}
We study the dynamical multistability of a solid-state single-atom laser
implemented in a quantum dot spin valve. The system is formed by a resonator
that  interacts with a two-level system in a dot in contact with two
ferromagnetic leads of antiparallel polarization. We show that a spin-polarized
current provides high-efficiency pumping leading to regimes of multistable
lasing, in which the Fock distribution of the oscillator displays a
multi-peaked distribution. The emergence of multistable lasing follows from
the breakdown of the usual rotating-wave approximation for the coherent
spin-resonator interaction which occurs at relatively weak couplings. The
multistability manifests itself directly in the  charge current flowing through
the dot, switching between distinct current levels corresponding to the
different states of oscillation.
\end{abstract}
%

\maketitle

\section{Introduction}
Quantum conductors coupled to localized harmonic resonators, such as microwave
photon cavities \cite{Liu:2015bha,Mi:2016ex,Viennot:2015ir,
PhysRevLett.115.046802,Stockklauser:2017bqa,Li:2018jr} or mechanical resonators
\cite{Naik:2006gf,Benyamini:2014eb,Okazaki:2016bh,Deng:2016iw} have become
commonly studied systems. They open the route to explore correlations between
charge transport and emitted radiation \cite{Lambert:2015} or induced
mechanical vibrations \cite{Parafilo:2016br}. Ultimately, these systems can
encode single-atom lasers which exhibit unique features compared to
conventional lasers such as absence of threshold, self-quenching, and
sub-Poissonian statistics \cite{Filipowicz:1986,Lugiato:1987ib,Mu:1992,
Rice:1994fj,Wang:1996fg}. Lasers where a cavity mode interacts with a stream
of excited atoms one at a time \cite{Filipowicz:1986,Lugiato:1987ib} can
display multistability \cite{Walther:2006da}, whereby two or more stable
amplitudes of oscillation coexist. Such behavior has also been predicted
to occur in solid-state analogues, such as single-electron transistors
\cite{PhysRevB.74.205336,Rodrigues:2007,Rodrigues:2007fv,
PhysRevLett.115.206802} and optomechanical systems \cite{Marquardt:2006,
Nation:2013}.
%
%
\begin{figure}[t!]
    \includegraphics[width=\columnwidth]{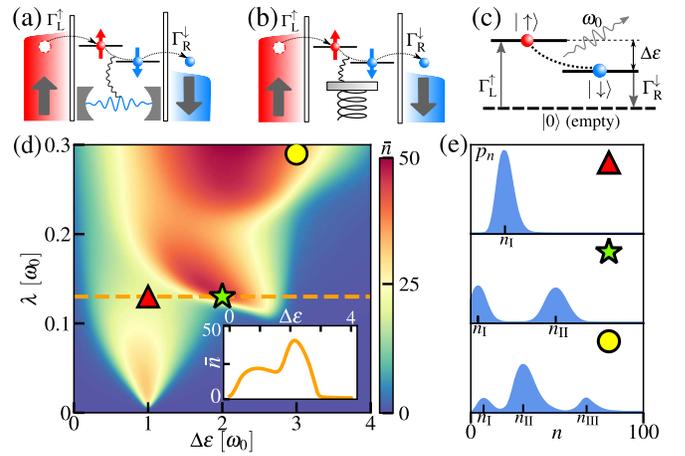}
    \caption{Model of a quantum-dot spin valve:
            (a) photon microwave cavity or (b)  nanomechanical resonator
            interacting with two spin levels of energy difference
            $\Delta \varepsilon$.
            Electron tunneling occurs at rates $\Gamma^\uparrow_\mathrm{L}$ and
            $ \Gamma^\downarrow_\mathrm{R} $  through ferromagnetic leads.
            (c) Energy diagram of the corresponding single-atom laser.
            (d) Average occupation of oscillator $\bar{n}$ as a function of
            spin energy splitting $\Delta\varepsilon$ and the spin-oscillator
            coupling strength $\lambda$ for fully polarized leads and
            $\Gamma^\uparrow_\mathrm{L} = \Gamma^\downarrow_\mathrm{R} =
            0.1\omega_0$, $Q = 10^3$ (inset: line at $\lambda = 0.13\omega_0$).
            (e) Steady-state Fock distributions $p_n$ at three different points
            (triangle, star and circle) marked in (d),
            with maxima at $n_\mathrm{I},n_\mathrm{II}$ and $n_\mathrm{III}$.
            }
    \label{fig:1}
\end{figure}


Single-atom lasers have been realized in cavity quantum electrodynamics (QED)
\cite{McKeever:2003gw,Walther:2006da}, in circuit QED \cite{Astafiev:2007gd}
and in hybrid systems with double quantum dots coupled to
microwave cavities \cite{Liu17}. Quantum dots are natural candidates for
exploring the rich physics of single-atom lasing, given their tunability and
versatility \cite{Childress:2004kt,Jin:2011gs,Liu:2014ge,Liu:2015bha,
Brandes:2003gu,Bergenfeldt:2013jz}. Theoretical works analyzed the possibility
of lasing in open quantum dots \cite{Childress:2004kt,Jin:2011gs}
and a number of successful experiments \cite{Liu:2014ge,PhysRevLett.114.196802,
Liu:2015bha,Liu17} reported lasing in double-quantum-dot systems. A single-atom
laser using spin-polarized current in spin-valve quantum dots has also been
proposed \cite{Khaetskii:2013}.

In this work, we show that a spin-valve quantum dot laser can display a rich
range of multistable dynamics. The emergence of multistability turns out to be
closely linked to the breakdown of the rotating-wave approximation (RWA), even
though it occurs for relatively weak dot-oscillator couplings. This is in
contrast to well-studied quantum optical systems which also display
multistability, such as the micromaser \cite{Walther:2006da,scully1997quantum}.
The spin-valve system therefore provides a very promising platform, not just
for studying unconventional laser-like dynamics in hybrid systems, but also
for investigating coherent spin-oscillator interactions beyond the RWA without
the requirement for ultrastrong couplings \cite{FornDiaz:2018,Kockum:2019ky}.
The spin-oscillator model we consider is depicted in Figs.~\ref{fig:1}(a)-\ref{fig:1}(c);
it comprises two levels of an electron spin with energy difference
$\Delta \varepsilon$ within a quantum dot embedded between ferromagnetic
contacts of opposite polarization. The spin interacts with a local resonator
of frequency $\omega_0$ which can be a microwave photon cavity,
Fig.~\ref{fig:1}(a), or a mechanical mode, Fig.~\ref{fig:1}(b).
Assuming strong Coulomb repulsion forbids double occupation in the dot,
the spin levels behave as a spin-$1/2$ interacting with the oscillator with
coupling strength $\lambda$. For a single resonator mode with large quality
factor and negligible relaxation rates for other (non-emitting) decay channels,
lasing is achieved, as illustrated in Fig.~\ref{fig:1}(d), as a function of
$\Delta \varepsilon$ and $\lambda$. Remarkably, regimes of bi- and
multistability are readily found where two or more states of large amplitude
of oscillation coexist, leading to corresponding maxima in the Fock
distribution of the resonator, as illustrated in Fig.~\ref{fig:1}(e). We show
that bistability can be achieved with experimentally accessible parameters
and detected using simple measurements of the average current flowing through
the dot.

This paper is organized as follows. In Sec. II, we introduce the model
Hamiltonian and the master-equation formalism. Section III describes 
the single-atom laser properties of the model within the RWA, while in Sec. IV, we
show how multistability emerges beyond the RWA. In Sec. V, we prove how the multistable dynamics can be detected through current
measurements, while Sec. VI is devoted to the experimental feasibility study 
of the system. 
Finally, we draw our conclusions in Sec. VII.
%
%

\section{Model Hamiltonian and Master Equation}
The dot-resonator system is described by the Rabi model
Hamiltonian $(\hbar=1)$
%
\begin{equation}
    \label{eq:system_hamiltonian}
    \hat{H}= \frac{\Delta\varepsilon}{2}\hat{\sigma}_z
    +
    \omega_0 \hat{b}^{\dagger} \hat{b}
    +
    \lambda(\hat{\sigma}_+ + \hat{\sigma}_-) (\hat{b}+ \hat{b}^{\dagger}),
\end{equation}
%
with  $\hat{b} , \hat{b}^{\dagger}$ the annihilation and creation
operators of the oscillator, $\hat{\sigma}_{\pm} = (\hat{\sigma}_{x} \pm i
\hat{\sigma}_{y})/2$ and $\hat{\sigma}_x,\hat{\sigma}_y,\hat{\sigma}_z$
Pauli spin operators associated to the two spin levels of the dot, polarized
in the $z$-direction, and with a transverse interaction with the oscillator
via $ \hat{\sigma}_x$.

In the limiting case of fully spin-polarized leads, the left contact fills the
spin-up level whereas spin-down electrons escape to the right, see
Fig.~\ref{fig:1}. The coherent interaction with the oscillator provides a
spin-flipping mechanism allowing an (inelastic) current to flow through the
dot accompanied by energy release into the oscillator: each electron passing
through the dot emits one quantum
of oscillation. However, the perfect correspondence between creation of quanta
and flow of current is broken if there is intrinsic spin relaxation in the dot,
or if the polarization in the leads is incomplete (so electrons can tunnel in
and out from both spin levels). When the lead polarizations are $P_{\nu}$,
with $\nu={\mathrm{L}},{\mathrm{R}}$, the spin-dependent tunneling rates are
given by $\Gamma_{\nu}^{\sigma} = \Gamma_{\nu} (1 + \sigma P_{\nu})/2$
for spin index $\sigma = \uparrow / \downarrow = +/-$. For simplicity, we
assume throughout symmetric and opposite polarization, i.e., $P_\mathrm{R} =
-P_\mathrm{L} = P$, with $0<P\leq 1$.

We focus on the regime $\Gamma_{\nu}^{\sigma} \ll  eV$, with
$V$ the bias voltage and $e$ the electron charge. Notice that the strong 
coupling limit is not necessary in our model since we can have $\lambda \ll 
\Gamma_\nu^\sigma$. For large bias voltage the average energy of the two spin 
levels is 
well
inside the bias window, and transport from right to left is blocked. In this
regime the dynamics is captured by a Markovian master equation in Lindblad form
for the  density matrix $\hat{\rho}$ of the coupled dot-resonator system
\cite{breuer2002theory,Timm:2008cl,CohenTannoudji:2012wp}. Tracing out the 
leads, and assuming local dissipation within each subsystem (dot and 
oscillator),  the master
equation at zero-temperature reads
%
\begin{equation}
    \label{eq:masterequation}
    \dot{\hat{\rho}} =
    -i[\hat{H}, \hat{\rho}]
    + \!\!\!\!  \sum_{\sigma=\uparrow,\downarrow}\
    [\Gamma^\sigma_\mathrm{L}\mathcal{D}(\hat{F}^\dagger_\sigma)\hat{\rho}
    + \Gamma^\sigma_\mathrm{R}\mathcal{D}(\hat{F}_\sigma)\hat{\rho}]
    + \kappa \mathcal{D}(\hat{b})\hat{\rho},
\end{equation}
%
where $\kappa$ is the oscillator damping rate (related to the quality factor
by $Q = \omega_0/\kappa$). We have denoted the Lindblad dissipator with
$\mathcal{D}(\hat{x})\hat{\rho} = \hat{x}\hat{\rho} \hat{x}^\dagger -
( \hat{x}^\dagger \hat{x} \hat{\rho} +
\hat{\rho}   \hat{x}^\dagger \hat{x}) /2$. The operators
\mbox{$\hat{F}_\sigma  = (1 - \hat{n}_{-{\sigma}}) \hat{d}_\sigma$} and
$F^\dagger_\sigma$ describe incoherent electron tunneling with the constraint
of vanishing double occupation. $\hat{d}_\sigma$ is the fermionic operator
which annihilates an electron of spin $\sigma$ in the dot and $\hat{n}_\sigma$
is the corresponding number operator.  The mappings $\hat{\sigma}_z =
\hat{n}_\uparrow - \hat{n}_\downarrow$ and $\hat{\sigma}_x =
\hat{d}^\dagger_\uparrow \hat{d}_\downarrow + \hat{d}^\dagger_\downarrow
\hat{d}_\uparrow$ hold in Hamiltonian~\eqref{eq:system_hamiltonian}.
The derivation of Eq.~\eqref{eq:masterequation} is given in 
Appendix~\ref{app:mastereq}. The steady-state solution of 
Eq.~\eqref{eq:masterequation} is found
numerically using the Python package QuTiP \cite{QUTIP,QUTIP2}.
%
%
%
%
\section{Standard single-atom laser and RWA}
\begin{figure}[t]
    \includegraphics[width=\columnwidth]{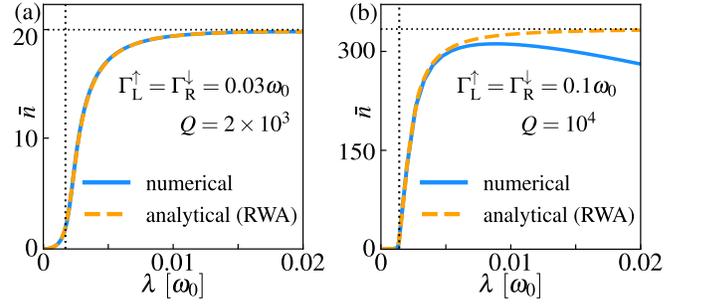}
    \caption{Average occupation $\bar{n}$ of the oscillator as a function of
             $\lambda$, for fully polarized leads at $\Delta\varepsilon =
             \omega_0$. Solid lines are numerical calculations, dashed lines
             are the analytical results from the Fock distribution $p_n$
             obtained within the RWA. The vertical and horizontal dotted lines
             are the threshold $\lambda_\mathrm{thr}$ and saturation number
             $A_s^2$, respectively, predicted by the semiclassical equations
             in RWA.
            }
    \label{fig:2}
\end{figure}

For weak spin-oscillator coupling,
the rotating-wave approximation (RWA) is expected to be valid and
$\hat{H}$ is approximated by
%
\begin{equation}
\label{eq:RWA_hamiltonian}
 \hat{H}^\mathrm{RWA} =
\frac{\Delta \varepsilon}{2} \hat{\sigma}_z
+
\omega_0 \hat{b}^{\dagger} \hat{b}
+
\lambda (\hat{\sigma}_{+} \hat{b} + \hat{\sigma}_{-} \hat{b}^{\dagger}).
\end{equation}
Using Eq.~\eqref{eq:RWA_hamiltonian} in Eq.~\eqref{eq:masterequation}
we recover well-known approximate analytical solutions for the oscillator Fock
probability distribution $p_n$ of a three-level single-atom laser
\cite{scully1997quantum}, see Appendix~\ref{app:RWA-1}. The average Fock 
number $\bar{n}= \langle
\hat{b}^\dagger \hat{b} \rangle$ calculated numerically coincides with the
analytical results in Fig.~\ref{fig:2}(a): incoherent pumping by electron
tunneling establishes a spin population inversion
leading to lasing.
%
%
%
%
%
By combining the RWA with a semiclassical approximation \cite{Mu:1992},
the operator $\hat{b}$ is replaced by  its time-dependent, classical
expectation value $\alpha(t)$, assuming quantum fluctuations  are negligible
(viz., above the lasing threshold). The spin is still described quantum
mechanically by a density matrix with $\rho_\uparrow(t)$ and
$\rho_\downarrow(t)$ the diagonal elements and $\rho_{\uparrow\downarrow}(t)$
the off-diagonal element. The dot occupation probability is
$p_1= \rho_{\uparrow}+\rho_{\downarrow}$ whereas
$S_z= \rho_{\uparrow}-\rho_{\downarrow}$ is the spin polarization.
Moving to a rotating frame with
$\alpha(t) \rightarrow \tilde{\alpha}(t)e^{-i\omega_0 t},
\rho_{\uparrow\downarrow}(t) \rightarrow
\tilde{\rho}_{\uparrow\downarrow}(t)e^{-i\Delta\varepsilon t}$,
we obtain a set of nonlinear equations for $\tilde{\alpha}(t)$, $p_1(t)$
and the spin vector $\vec{S}(t) = [S_x(t), S_y(t), S_z(t)]^T$ with
$ \tilde{\rho}_{\uparrow\downarrow}(t) = [S_x(t) - i S_y(t)]/2$, derived
in Appendix~\ref{app:RWA-2}.
Within this framework, lasing is equivalent to self-sustained oscillations:
the relaxation dynamics for the amplitude  $\left|\tilde{\alpha}\right|=A$
is given by
%
%
\begin{equation}
\label{eq:amplitude_equation}
 \dot{A}   = - [\kappa  + \gamma_\mathrm{rw}(A)]\frac{A}{2}
\end{equation} 
%
with the
effective, negative nonlinear damping
%
\begin{equation}
\label{eq:gammaeff_RWA}
\gamma_\mathrm{rw}  (A)  = - \frac{  \lambda^2\Gamma_\mathrm{eff}} {
\lambda^2 A^2 + \Gamma_\mathrm{eff}   \Gamma^{\downarrow}_\mathrm{R}/4}
\end{equation}
%
where $ \Gamma_\mathrm{eff} =
\Gamma^{\uparrow}_\mathrm{L} \Gamma^{\downarrow}_\mathrm{R} /
(2\Gamma^{\uparrow}_\mathrm{L} + \Gamma^{\downarrow}_\mathrm{R})$.
Equation~\eqref{eq:amplitude_equation} predicts a stable steady-state solution 
of finite $A$
above a threshold coupling $\lambda_\mathrm{thr}$.
For fully polarized leads ($P=1$) and on resonance
($\Delta\varepsilon = \omega_0$), one
obtains $\lambda_\mathrm{thr}^2 =
\Gamma^{\downarrow}_{\mathrm{R}}\omega_0/(4Q)$
and for $\lambda \gg \lambda_\mathrm{thr}$ the amplitude saturates
to $A_s = \sqrt{ \Gamma_\mathrm{eff} Q/\omega_0}$.
Semiclassical predictions for the saturation and threshold are shown
as straight lines in Fig.~\ref{fig:2}. 

We conclude by observing that, for finite polarization, we have the weak
scaling $\lambda_\mathrm{thr}  \sim 1/\sqrt{P}$, $A_s \sim \sqrt{P}$, as
shown in Appendix~\ref{app:RWA}.
%
%
%
%
\section{Multistability beyond RWA}
\begin{figure}[b]
	\includegraphics[width=\columnwidth]{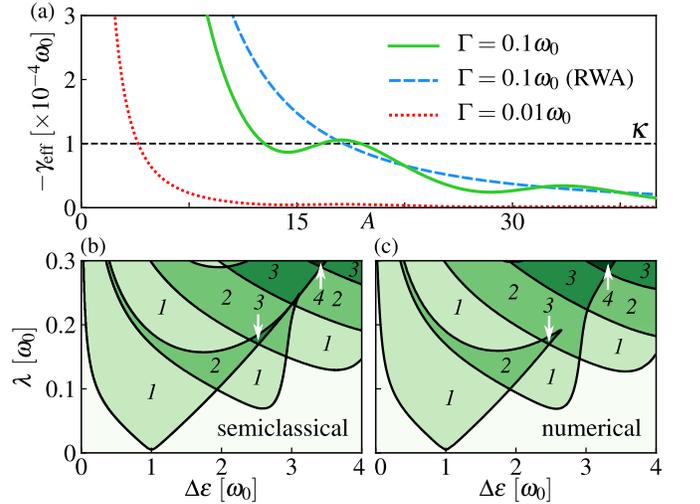}
	\caption{(a) Effective negative nonlinear damping
		$\gamma_{\mathrm{eff}}(A)$ for the oscillator,
		as given by Eq.~\eqref{eq:gammaeff} on resonance
		$\Delta\varepsilon = \omega_0$ and
		for $\lambda = 0.055\omega_0$.
		Intersections with the horizontal dashed line
		at $\kappa = 10^{-4}\omega_0$  indicate limit cycles for the
		amplitude.
		(b)  Multistability diagram obtained with the semiclassical
		approximation, showing the number of stable limit cycles
		(italic numbers) as a function of $\Delta \varepsilon$ and
		$\lambda$.
		(c) Multistability diagram obtained from the numerical
		solution of the Lindblad equation (number of peaks in the Fock
		distribution).
		Parameters: $Q = 10^3,\ P=1,\  \Gamma = 0.05 \omega_0$.
	}
	\label{fig:3}
\end{figure}

The RWA predicts that the saturation amplitude
$A_s$ should simply increase with increasing $Q$ and $\Gamma_\mathrm{eff}$,
without any other changes developing. However, numerical calculations show
that the average Fock occupation ($\bar{n}$) no longer saturates and
instead drops with increasing $\lambda$, see Fig.~\ref{fig:2}(b).
This breakdown in the RWA occurs when the  Rabi oscillation frequency
of the spin (which is proportional to $\lambda A$) approaches  $\omega_0$.
In fact, for large enough Rabi frequencies, the Fock distribution (obtained
numerically) becomes \mbox{multi-peaked} with the highest peak close to the
amplitude predicted by the RWA, Fig.~\ref{fig:1}(e). This happens even at
finite detuning ($\Delta \varepsilon \neq \omega_0$) giving rise to the complex
behavior of $\bar{n}$ reported in Fig.~\ref{fig:1}(d).
By extending the semiclassical approach to analyze the behavior beyond RWA,
we show that the oscillator dynamics can possess two or more coexisting stable
limit cycles with different amplitudes. The resulting phase diagram of the bi-
and multistable regions agrees closely with the numerical results as shown
in Figs.~\ref{fig:3}(b) and \ref{fig:3}(c).

Focusing on the case $\Gamma^{\uparrow}_{\mathrm{L}} =
\Gamma^{\downarrow}_{\mathrm{R}}/2=\Gamma$
to simplify the discussion,  $p_1$ becomes irrelevant and we write again a
set of nonlinear equations for $\tilde{\alpha}(t)$ and  $\vec{S}(t)$ in the
rotating frame (see Appendix~\ref{app:beyondRWA} for details). In the regime  
$\kappa \ll \lambda, \Gamma, \omega_0$, the
oscillator amplitude $|\tilde{\alpha}| = A$ is a slow variable while its phase
is irrelevant and can be set to zero. Assuming constant $A$,
the equation for the spin vector when $\Delta \varepsilon = \omega_0$
and $P=1$ is
%
\begin{equation}
    \label{eq:system_beyond_RWA}
    \dot{\vec{S}}(t) = \Gamma\hat{u}_z - \Gamma\vec{S}(t) + \vec{S}(t)
    \times \vec{B}(t),
\end{equation}
with $\hat{u}_z $ unit vector in the $z$-direction and
\begin{eqnarray}
    \vec{B}(t) =
    \left(
    \begin{array}{c}
        B_x(t)  \\
        B_y(t)
    \end{array}
    \right)
    =
    2 \lambda A
    \left(
    \begin{array}{c}
        - 1 -\cos(2\omega_0 t)   \\
        \sin(2\omega_0 t)
    \end{array}
    \right).
    \label{eq:effective_field}
\end{eqnarray}
%
The behavior of the solutions of
Eqs.~\eqref{eq:system_beyond_RWA}-\eqref{eq:effective_field}
is similar to that seen in previous studies on circuit-QED systems
\cite{Rodrigues:2007fv} and is related to a phase-locking phenomenon
in which the Rabi frequency of the spin---determined by the oscillation
amplitude---seeks to be commensurate to the oscillator frequency
\cite{Marquardt:2006}. By writing Eq.~\eqref{eq:system_beyond_RWA} in Fourier
space, with 
$
S_k(t) = \sum_n S_k^{(n)} e^{2i n \omega_0 t},\ (
k = x, y, z),$
we obtain a recursion relation for the $A$-dependent
Fourier coefficient $S_z^{(n)}$ in
terms of $S_z^{(n\pm1)}$.
We can then calculate the amplitude-dependent effective negative nonlinear
damping  $\gamma_{\mathrm{eff}}$ acting on the oscillator due to the spin
dynamics,
%
\begin{equation}
    \label{eq:gammaeff}
    \gamma_{\mathrm{eff}}(A) = -\frac{2 \lambda^2}{\Gamma}
    \left[\frac{4\omega_0^2}{\Gamma^2+4\omega_0^2}S_z^{(0)}
    - \mathrm{Im} \left(\frac{2 \omega_0}{\Gamma + 2 i \omega_0}S_z^{(1)}
    \right)\right].
\end{equation}
%
We show the results for $\gamma_{\mathrm{eff}}$ on resonance in
Fig.~\ref{fig:3}(a). The monotonic RWA behavior
$\gamma_{\mathrm{eff}}\approx \gamma_\mathrm{rw}$ is recovered only at low
tunneling rate whereas the function $\gamma_{\mathrm{eff}}$ oscillates at
larger $\Gamma$ with maxima close to the points $\lambda A \sim  j \omega_0$
with $j$ integer. This nonmonotonic behavior leads to many (stable) limit
cycles determined by the intersections $\kappa=-\gamma_{\mathrm{eff}}(A)$
with a negative slope of $| \gamma_\mathrm{eff}(A) |$.
Equation~\eqref{eq:gammaeff} is readily generalized to the off-resonant case 
in Appendix~\ref{app:beyondRWA-2}
and we can extract the stable steady-state amplitudes to produce the predicted
multistability diagram, Fig.~\ref{fig:3}(b). We test the validity of the
semiclassical solution by finding numerically the steady-state Fock
distribution $p_n$ of the system through Eq.~\eqref{eq:masterequation} and
computing the number of distinct peaks in $p_n$ with $n>0$,
see Fig.~\ref{fig:3}(c).

The semiclassical method has the important advantage that it can be used to
calculate the onset of bi- and multistability at relatively weak coupling
strengths, $\lambda \sim 10^{-3} \omega_0$, and high quality factors,
$Q \sim 10^5$, which are most likely to be accessible experimentally
(as discussed below). For large $Q$, the average occupation number is too 
large to allow a full numerical solution 
of the master equation, since it requires a prohibitively 
large cutoff in the Fock state basis.
%
%
%
%
%
\section{Current jumps}
\begin{figure}[t]
    \includegraphics[width=\columnwidth]{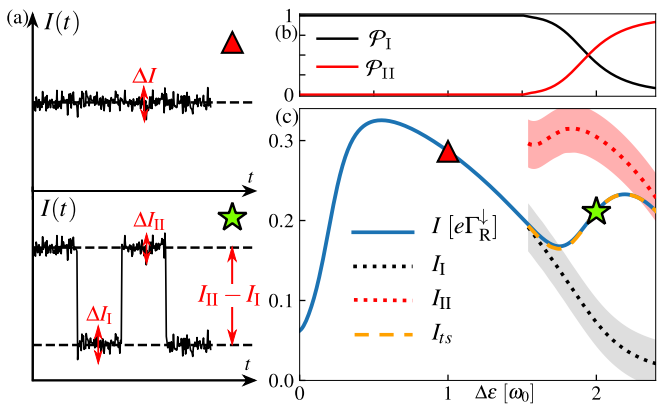}
    \caption{(a) Sketch of the behavior of the current in the lasing regime
             with a single stable oscillator amplitude $A_\mathrm{I}$ (upper)
             and in a bistable regime with distinct stable amplitudes
             $A_\mathrm{I}$ and $A_\mathrm{II}$ (lower).
             (b) Occupation probabilities $\mathcal{P}_\mathrm{I}$ and
             $\mathcal{P}_\mathrm{II}$ of the two states computed numerically.
             (c) Average current computed numerically (solid blue line) and 
             using
             the effective two-state model (dashed orange line). The upper 
             (red) and lower (black) dotted lines show the two current states 
             $I_\mathrm{I}$ and $I_\mathrm{II}$ along with their variance, 
             represented by the shaded areas.
             The parameters match those of the inset in Fig.~\ref{fig:1}(d).
            }
    \label{fig:4}
\end{figure}

Measurement of the dc-current through the dot
provides a simple way to detect lasing and bistability.
In the large bias limit the current is given  by
%
%
\begin{equation}
\label{eq:current}
I =  e(\Gamma_{\mathrm{R}}^\uparrow \rho_\uparrow +
\Gamma_{\mathrm{R}}^\downarrow \rho_\downarrow ).
\end{equation}
In the fully polarized case the total current is purely inelastic
(oscillator-assisted spin flips), $I = e \Gamma_{\mathrm{R}}^{\downarrow}
\rho_{\downarrow}$, and on resonance we have $\kappa \bar{n}   = I/e$,
as expected by energy conservation: the outgoing flux of quanta equals
the ingoing flux of energy into the oscillator.
For large oscillator occupation number (i.e.,\ large oscillation amplitudes
in the semiclassical framework), the average current is much larger than
its fluctuations and acts as a measure of the oscillator amplitude as
illustrated in Fig.~\ref{fig:4}(a).
%
%
%
%
The current also provides a simple way to detect the RWA breakdown and
the onset of bistable regime, as it can display telegraph dynamics.
For a well-developed bistability, the oscillator exists in a mixed state
containing two different limit cycles with well-separated amplitudes
$A_\mathrm{I}$ and $A_{\mathrm{II}}$ and associated probabilities
$\mathcal{P}_\mathrm{I}$ and $\mathcal{P}_\mathrm{II}$. The amplitude is
then expected to switch randomly between two well-defined plateaus when
$\mathcal{P}_\mathrm{I} \sim \mathcal{P}_\mathrm{II}$.
The close connection between current and oscillator amplitude suggests
that the telegraph dynamics will manifest itself in random switching between
plateaus of different average current associated with the different states of
oscillation \cite{Kirton:2013cm}, as sketched in Fig.~\ref{fig:4}(a). Such
behavior is also naturally implied by the semiclassical treatment in which,
for each stable amplitude solution $A_\mathrm{I}, A_\mathrm{II}$ of the
oscillator, one has a different solution for the average current,
$I_\mathrm{I}$ and $I_\mathrm{II}$.

Telegraph behavior of the current should be observable if:
(i)  $\mathcal{P}_\mathrm{I} \sim \mathcal{P}_\mathrm{II}$;
(ii)  the variance associated to each plateau is smaller than
the distance:
$\Delta I_\mathrm{I},\Delta I_\mathrm{II}\ll |I_\mathrm{I}-I_\mathrm{II}|$;
(iii) the lifetime of each plateau is sufficiently long to observe separated
jumps.
Under these conditions the
system is well described by an effective two-state model with
transition rates $W_{\mathrm{I} \rightarrow \mathrm{II}}$ and
$W_{\mathrm{II} \rightarrow \mathrm{I}}$. We test the appropriateness of the
two-state model by computing the probabilities  $\mathcal{P}_\mathrm{I}$,
$\mathcal{P}_\mathrm{II}$ (from the areas of the two peaks in the $p_n$
distribution) and by comparing the average dc-current calculated numerically
with the two-state expression 
%
\begin{equation}
\label{eq:current_ts}
I_{ts} =  \mathcal{P}_\mathrm{I} I_\mathrm{I} +
\mathcal{P}_\mathrm{II} I_\mathrm{II},
\end{equation}
%
%
see Fig.~\ref{fig:4}(b,c).
Here we also report the current variance for each plateau
$\Delta I_\mathrm{I}$ and $\Delta I_\mathrm{II}$
defined as
$\Delta I_\mathrm{I} = \sqrt{\Gamma S_\mathrm{I}}$
and
$\Delta I_\mathrm{II} = \sqrt{\Gamma S_\mathrm{II}}$,
with $S_\mathrm{I}$, $S_\mathrm{II}$
the current shot-noise.
Finally, we obtain the sum of the two rates
$W_{\mathrm{I} \rightarrow \mathrm{II}} +
W_{\mathrm{II} \rightarrow \mathrm{I}}$
by comparing the current shot-noise, calculated
numerically with the full counting statistics method, to the two-state formula 
(see Appendices~\ref{app:FCS}-\ref{app:TSM} and Refs.
\onlinecite{PhysRevB.70.205334,Flindt2005,PhysRevB.82.155407,blanter,
PhysRevB.85.125441} for details).
This sum agrees well with the real part of the smallest nonzero
eigenvalue of the system Liouvillian, verifying the applicability of
the two-state model and showing that the switching can
be extremely slow (orders of magnitude slower than the relaxation time
of the oscillator) \cite{PhysRevB.78.024513} as discussed in 
Appendix~\ref{app:TSM}.
%
%
%
\section{Experimental feasibility}
At finite polarization the total
current has an elastic contribution in addition to the inelastic one that
arises from the interaction with the oscillator. This leads to lower
efficiency, but lasing and multistability are still achievable. Importantly,
the inelastic current is still a substantial fraction of the total current
(larger than the noise) such that the current jumps are still clearly
observable. Using numerical calculations, we test that the results presented
so far are robust against effects of finite polarization.
Having in mind the case of mechanical oscillators, we also include the
effects of finite temperature $T$
(namely, $n_\mathrm{B} > 0 $  with
$n_\mathrm{B} = [\exp (\omega_0 / T )- 1]^{-1}$
the thermal bosonic occupation number and $k_\mathrm{B} = 1$) and
intrinsic nonlinearity
\cite{Meerwaldt2012}.
The multistability is  preserved in a substantial range of parameters far from
the ideal case (e.g., $P = 0.5$) including finite internal spin relaxation of
the dot, which plays a role similar to finite polarization: impinging spin-up
electrons can decay into the spin-down level and pass through the dot without
quanta emission. When the spin relaxation rate $\gamma_\mathrm{sr}$ is smaller
than the tunneling rates $\Gamma$ and the Rabi frequency $\lambda A$,
lasing and multistability remain unperturbed. Several examples of the 
behaviour of the results including finite temperature, finite polarization, 
spin relaxation and Duffing nonlinearity are shown in 
Appendix~\ref{app:nonideal}.

Spin-valve-based carbon nanotube quantum dots (CNTQDs) provide a promising way
of implementing the model system we have investigated. CNTQDs can achieve
high spin polarization of injected electrons \cite{Sahoo:2005dw} and small spin
relaxation rate \cite{Rice:2013ep,Churchill:2009fy,Viennot:2015ir}.
Furthermore, suspended nanotubes act as electromechanical systems with
vibrational modes of huge quality factor \cite{Moser:2014jea}. Spin-vibration
interaction in suspended CNTQDs has been investigated theoretically in
spin-valve setups
\cite{Palyi:2012,PhysRevLett.113.047201,Stadler:2015}.
For $Q = 10^6$, $P=0.5$, $\Gamma_\mathrm{L} = \Gamma_\mathrm{R}= 0.05\omega_0$,
we estimate a threshold $\lambda_\mathrm{thr} \approx 1.6 \times 10^{-4}
\omega_0$ which is well below the expected interaction strength
$\lambda \approx 1\ \text{MHz}$ for a typical resonance frequency
$\omega_0/2\pi = 100\ \text{MHz}$ \cite{PhysRevLett.113.047201,Stadler:2015}.

Realizations based on spin valves coupled to microwave cavities should also be
possible. Reliable coupling of CNTQDs with superconducting microwave cavities
has been demonstrated \cite{Ranjan:2015jk}. Spin-photon interactions have also
been implemented in quantum dots with ferromagnetic leads \cite{Viennot:2015ir}
and, more recently, in silicon double dots embedded in magnetic
nanostructures \cite{Mi:2018ip}.
%
%
%
\section{Conclusions}
We have analyzed a model quantum dot spin valve which
forms an unconventional single-atom laser: a spin-polarized current pumps the
motion of a resonator coupled to the dot very efficiently, allowing access to
novel regimes of multistable lasing. We show that multistability develops when
the dot-resonator interaction is no longer captured by the conventional
RWA---which is expected to occur for the relatively weak couplings achievable
with current devices---because large amplitude motion of the resonator enhances
the effective coupling strength. This type of system provides an alternative
route for investigating coherent dynamics beyond the RWA without the need for
ultrastrong couplings. Our work raises a range of interesting questions about
the extent to which the multistable lasing dynamics can be controlled
and exploited, e.g., in nonlinear amplifiers or force sensing devices.
%
%
%
%
\begin{acknowledgments}
    We thank Mark Dykman, Christian
    Flindt and Fabio Pistolesi for useful discussions.
    This research was supported by the German Excellence
    Initiative through the Zukunftskolleg and the Deutsche
    Forschungsgemeinschaft (DFG) through the SFB 767.
\end{acknowledgments}
%
%
%
\appendix
%
%
%
\section{Derivation of the master  equation}
\label{app:mastereq}
In this Appendix we derive Eq.~\eqref{eq:masterequation} and 
discuss critically its validity regime. We start from the  model Hamiltonian  
that describes a quantum dot with 
spin-dependent levels, between two lateral 
leads, and coupled to an harmonic oscillator damped through a bosonic thermal
bath ($\hbar = 1$):
\begin{align}
	\hat{H}_\mathrm{tot} &= \hat{H} + 
	\hat{H}_\mathrm{leads} +
	\hat{H}_\mathrm{tun} + 
	\hat{H}_\mathrm{bath} +
	\hat{H}_\mathrm{osc-bath},
	\label{eq:total_hamiltonian} 
\end{align}
with
\begin{align}
	\hat{H} &= \varepsilon_0 (\hat{n}_\uparrow + \hat{n}_\downarrow) +
	\frac{\Delta\varepsilon}{2}(\hat{n}_\uparrow - \hat{n}_\downarrow) 
	+ U \hat{n}_\uparrow \hat{n}_\downarrow \nonumber \\
	&\,\,+ \omega_0 \hat{b}^\dagger \hat{b}
	+ \lambda (\hat{d}^\dagger_\uparrow \hat{d}_\downarrow 
	+ \hat{d}^\dagger_\downarrow \hat{d}_\uparrow) (\hat{b} + 
	\hat{b}^\dagger),  
	\label{eq:H_S} \\
	\hat{H}_\mathrm{leads}&=
	\sum_{\nu=\mathrm{L \,,R}}  \sum_{k \sigma} 
	(\epsilon_{\nu k \sigma}-\mu_\nu) \hat{c}^{\dagger}_{\nu k \sigma} 
	\hat{c}_{\nu k \sigma}^{\phantom{g}}, \\
	\hat{H}_\mathrm{tun} &=
	\sum_{\nu=\mathrm{L \,,R}}  \sum_{k \sigma} V_{\nu \sigma} 
	\left( \hat{c}_{\nu k \sigma}^{\phantom{g}}  \hat{d}^\dagger_{\sigma} + 
	\mathrm{H.c.} \right), \\
	\hat{H}_\mathrm{osc-bath} &= \hat{b} \hat{B} + \mathrm{H.c.}
\end{align}
We have labeled with $\varepsilon_0$ the average energy of the 
two levels in the dot and $\Delta \varepsilon$ their energy separation. 
The Coulomb interaction is taken into account via the repulsive energy 
$U>0$ for the doubly-occupied state.
$\hat{H}_\mathrm{leads}$ corresponds to the leads, viz., two Fermi gases, 
with  $\hat{c}_{\nu k \sigma}$ the annihilation operator for a level of 
energy $\epsilon_{\nu k \sigma}$ on the $\nu$ lead 
kept at  chemical potential $\mu_\nu$.
The coupling between the leads and the dot is realized through the 
tunneling Hamiltonian $\hat{H}_\mathrm{tun} $,
with $V_{\nu \sigma}$ the tunneling amplitudes. 
Finally, the oscillator is linearly coupled to a bosonic bath (described by 
$\hat{H}_\mathrm{bath}$) through the operator of the bath $\hat{B}$.
\subsection{Born-Markov master equation}
\label{sec:IA}
We identify our system as the dot coupled to the oscillator,
evolving coherently under Hamiltonian \eqref{eq:H_S}, and we seek for the 
Markovian master equation describing the evolution of the system density 
matrix 
$\hat{\rho}$, using the standard open systems approach 
\cite{breuer2002theory,CohenTannoudji:2012wp}. The 
external 
environment
is described by $\hat{H}_E = \hat{H}_\mathrm{leads} + \hat{H}_\mathrm{bath}$,
interacting with the system through $\hat{H}_\mathrm{int} = 
\hat{H}_\mathrm{tun} + \hat{H}_\mathrm{osc-bath}$. In the interaction picture
with respect to $\hat{H} + \hat{H}_E$, the exact equation for the total 
density matrix $\hat{\rho}_{\mathrm{tot}}$ is
\begin{align}
	\label{eq:rho_total}
	\dot{\hat{\rho}}_{\mathrm{tot},I} (t)
	=  &- i\left[ \hat{H}_{\mathrm{int}, I}(t) , 
	\hat{\rho}_{\mathrm{tot},I}(t_0) 
	\right] \nonumber \\
	&- \int_{t_0}^{t} \!\!\!\! \mathrm{d}t' \,\, 
	\left[
	\hat{H}_{\mathrm{int},I}(t), \left[\hat{H}_{\mathrm{int},I}(t'),
	\hat{\rho}_{\mathrm{tot},I}(t') 
	\right] 
	\right],
\end{align}
where the subscript $I$ refers to the interaction picture. At this point,
a number of assumptions are in order. (i) We assume that the interaction with
the leads and the bath is turned on at some initial time $t_0$. Up to 
this instant, the total density matrix is factorized, 
$\hat{\rho}_\mathrm{tot}(t_0) = \hat{\rho}(t_0) 
\hat{\rho}_\mathrm{leads}(t_0) \hat{\rho}_\mathrm{bath}(t_0)$ (the tensor 
product
is implied); the 
reservoirs are at separate thermal equilibria (the leads can have different 
chemical potentials). (ii) The internal correlations in the environments decay 
on a timescale which is much shorter than the timescale of interaction between 
the 
dot and the leads (given in the interaction picture by the inverse of the 
average tunneling amplitude $\overline{V_{\nu \sigma}}$) and between the 
oscillator and its
bath. This follows from the assumption that the reservoirs are weakly coupled 
to the system and are very large, reaching thermal equilibrium very fast: 
their state is weakly affected by 
the interaction with the system, such that one can 
replace $\hat{\rho}_{\mathrm{tot},I}(t')$ with $\hat{\rho}_I(t') 
\hat{\rho}_{\mathrm{leads},I}(t_0)
\hat{\rho}_{\mathrm{bath},I}(t_0)$ in the integral. This weak-coupling 
approximation is 
commonly referred to as Born approximation 
\cite{CohenTannoudji:2012wp}. (iii) The existence of a timescale
separation allows us to make Eq.~\eqref{eq:rho_total} local in time, such that 
the
evolution of $\hat{\rho}$ at time $t$ only depends on $\hat{\rho}$ at the same
instant (Markov approximation). By finally transforming back to the 
Schr\"odinger picture, we write the Wangsness-Bloch-Redfield master equation 
\cite{Timm:2008cl}:
\begin{alignat}{3}
	\label{eq:WBR}
	\dot{\hat{\rho}}(t) &= &&-\!i [\hat{H}, \hat{\rho}(t)] \nonumber \\& 
	&&-\!\! 
	\int_0^{\infty}\!\!\!\! \mathrm{d}\tau \Tr_E \{[\hat{H}_\mathrm{int}, 
	[\hat{H}_\mathrm{int}(-\tau), \hat{\rho}(t)
	\hat{\rho}_\mathrm{leads} (t_0) \hat{\rho}_\mathrm{bath} (t_0)] ]\} 
	\nonumber\\
	&\equiv &&\mathcal{L} \hat{\rho}(t),
\end{alignat}
where we introduced the total Liouvillian superoperator $\mathcal{L}$.
Its action on $\hat{\rho}$ can be decomposed into the sum of the coherent part
$- i [\hat{H}, \hat{\rho}]$ and the 
dissipative part $\mathcal{L}_\mathrm{leads}\hat{\rho} +
\mathcal{L}_\mathrm{bath}\hat{\rho}$.
The decomposition is possible because the leads
and the bath are uncorrelated reservoirs.
\subsection{Large bias voltage and strong Coulomb repulsion limit}
We consider a bias voltage $V$ applied symmetrically to the leads, such that
$\mu_\mathrm{L} = eV/2$ and $\mu_\mathrm{R} = -eV/2$. Next, we assume the 
limit of large voltage bias. Thus, the 
Fermi functions
$f_\nu (\epsilon) = \{\exp[(\epsilon - \mu_\nu)/T] + 1\}^{-1}$ for the leads
($k_\mathrm{B} = 1$ and $T$ is the temperature) can be approximated to be 
$f_\mathrm{L} \approx 1$ and
$f_\mathrm{R} \approx 0$, independent on the energy. All energy levels
of the system lie inside the bias window, and electron transport from right to
left is blocked. Computing the time integrals in Eq.~\eqref{eq:WBR} in
the large bias limit, we can write the dissipator for the leads as
\begin{equation}
	\mathcal{L}_\mathrm{leads} \hat{\rho} = 
	\sum_{\sigma = \uparrow, \downarrow} \left[
	\Gamma^\sigma_\mathrm{L} \mathcal{D}(\hat{d}^\dagger_\sigma) \hat{\rho} +
	\Gamma^\sigma_\mathrm{R} \mathcal{D}(\hat{d}_\sigma) \hat{\rho} \right]. 
	\label{eq:leads_dissipator}
\end{equation}
%
%
The bare tunneling rates are given by $\Gamma^\sigma_\nu = 2\pi 
|V_{\nu\sigma}|^2 \rho_{\nu \sigma}$, with $\rho_{\nu\sigma}$ the
spin-$\sigma$ density of states at the Fermi level of lead $\nu$. We
have made here the wide-band approximation, such that the spectral densities
of the dot-lead couplings are energy-independent. At low temperature, the
correlation functions of the leads decay on a timescale
$\tau_\mathrm{leads} \approx \hbar/eV$ (we restore the Planck's constant for
the moment) \cite{Timm:2008cl}, and become indeed the smallest timescale
required in assumptions (ii)-(iii) of Section~\ref{sec:IA} in the large
bias limit. 
The leads are ferromagnetic, with a finite polarization $P_\nu$ for lead $\nu$.
We can write the tunneling rates as
$\Gamma^\sigma_\nu = \Gamma_\nu (1 +\sigma P_\nu)/2$. For symmetric and 
opposite polarizaion $P$ the rates read:
\begin{align}
	&\Gamma_{\mathrm{L}}^\uparrow = \Gamma_\mathrm{L} \left( \frac{1 + P}{2} 
	\right),
	\quad
	\Gamma_{\mathrm{L}}^\downarrow = \Gamma_\mathrm{L} \left( \frac{1 - P}{2} 
	\right),
	\nonumber \\
	&\Gamma_{\mathrm{R}}^\uparrow = \Gamma_\mathrm{R} \left( \frac{1 - P}{2} 
	\right),
	\quad
	\Gamma_{\mathrm{R}}^\downarrow = \Gamma_\mathrm{R} \left( \frac{1 + P}{2} 
	\right).
	\label{eq:tunneling_rates}
\end{align}
We now assume that the Coulomb repulsion $U$ inside the quantum dot becomes 
the 
largest energy scale in the system, i.e., one has also $U \gg eV$. The
doubly-occupied state is away from the bias window and cannot be even thermally
populated at finite temperature $T$. In this limit, the population of 
the doubly-occupied state and the coherences involving this state are 
constrained
to vanish by replacing the
dot operator $\hat{d}_\sigma$ with $\hat{F}_\sigma = (1 - \hat{n}_{-\sigma})
\hat{d}_\sigma$, together with its complex conjugate, in 
Eq.~\eqref{eq:leads_dissipator}. Simultaneously, one can remove the Coulomb 
term from Hamiltonian \eqref{eq:H_S}. The dot is either empty
or singly-occupied due to the incoherent single-electron tunneling events. 

To obtain Eq.~\eqref{eq:masterequation}, we assume that the dissipation for 
the harmonic 
oscillator (described by $\mathcal{L}_\mathrm{bath}$) can be added locally in 
the standard way,
assuming that the quality factor $Q$ is very large (the oscillator is very 
weakly
coupled to its bath, and it is extremely underdamped) 
\cite{scully1997quantum,breuer2002theory}. Equation~\eqref{eq:WBR} becomes 
finally
\begin{align}
	\dot{\hat{\rho}} = &-i[\hat{H}, \hat{\rho}] + 
	\sum_{\sigma = \uparrow, \downarrow} \left[
	\Gamma^\sigma_\mathrm{L} \mathcal{D}(\hat{F}^\dagger_\sigma) \hat{\rho} +
	\Gamma^\sigma_\mathrm{R} \mathcal{D}(\hat{F}_\sigma) \hat{\rho} \right] 
	\nonumber \\
	&+\kappa (1 + n_\mathrm{B}) \mathcal{D} (\hat{b}) + \kappa  n_\mathrm{B} 
	\mathcal{D} (\hat{b}^\dagger),
	\label{eq:master_equation}
\end{align}
with the intrinsic damping of the resonator, $\kappa = \omega_0/Q$, and the
average number of excitations in the thermal bath at frequency $\omega_0$ 
and temperature $T$, given by
$n_\mathrm{B} = [\exp(\omega_0/ T) - 1]^{-1}$. Setting $n_\mathrm{B} = 0$ 
gives the 
zero-temperature limit illustrated by Eq.~\eqref{eq:masterequation}.

We conclude by explaining the equivalence between Eqs.~\eqref{eq:H_S} and 
\eqref{eq:system_hamiltonian}. The coherent dynamics of the system does 
not involve 
the empty and the doubly-occupied state. The dot's Hilbert space is thus 
reduced to that of a two-level system. This allows us to map the dot operators 
to the 
Pauli algebra through $\hat{n}_\uparrow - \hat{n}_\downarrow 
\rightarrow \hat{\sigma}_z$, and $\hat{d}^\dagger_\uparrow \hat{d}_\downarrow 
+ \hat{d}^\dagger_\downarrow \hat{d}_\uparrow \rightarrow \hat{\sigma}_x$, 
after
projecting out the irrelevant states. In the formal solution of 
Eq.~\eqref{eq:master_equation} the empty state must be taken into account. 
Finally, the average energy level $\varepsilon_0$ of the quantum dot is 
irrelevant
in the open dynamics and can be disregarded, because we work in the large bias 
limit.
%
%
%
%
%
%
\section{Single-atom laser within the RWA}
\label{app:RWA}
\subsection{Analytical solution for the steady-state Fock distribution}
\label{app:RWA-1}
In the rotating-wave approximation (RWA) we can obtain an analytical 
expression for the steady-state 
Fock distribution $p_n$ of the harmonic oscillator and show that it 
corresponds to a lasing state.
Starting from the Eq.~\eqref{eq:master_equation},  we replace
the system Hamiltonian with Eq.~\eqref{eq:RWA_hamiltonian}.
We discuss here the resonant case, $\Delta \varepsilon = \omega_0$.
Following standard textbooks \cite{scully1997quantum} we assume a large quality
factor for the oscillator and derive the equation for the steady-state 
Fock distribution $p_n$ in recursive form:
\begin{equation}
	\label{eq:recursive_equation_pn}
	\left[ \frac{n \kappa (\lambda/\lambda_{\mathrm{thr}})^2}{1 + 
	\frac{n}{A^2_s} (\lambda/\lambda_{\mathrm{thr}})^2} + \kappa n_\mathrm{B} 
	n \right] p_{n-1} = \kappa (1 + n_\mathrm{B}) p_n,
\end{equation}
with
\begin{equation}
	\label{eq:nsat_threshold_p}
	A^2_s = \frac{\Gamma_{\mathrm{L}} \Gamma_{\mathrm{R}} P}{ (2 
	\Gamma_{\mathrm{L}}+\Gamma_{\mathrm{R}} )\kappa},
	\,\,\, \lambda_\mathrm{thr} = \sqrt{\frac{\Gamma_{\mathrm{R}} \kappa}{4} 
	\!\!\left[ \frac{\!2 \Gamma_{\mathrm{L}}(1\! +\! P^2) + 
	\Gamma_{\mathrm{R}}(1\! -\! P^2)}{4 \Gamma_{\mathrm{L}} P}\!\right]}.
\end{equation}
$A_s^2$ is the saturation number, while $\lambda_\mathrm{thr}$ is the 
threshold coupling. The solution to Eq.~\eqref{eq:recursive_equation_pn} can 
be written as
\begin{equation}
	\label{eq:steadystate_distribution}
	p_n = p_0 \frac{\mathscr{N}_n}{\mathscr{D}_n} 
	\left(\frac{n_\mathrm{B}}{n_\mathrm{B}+1} \right)^n.
\end{equation}
We introduced the Pochhammer symbol, $a_n = a(a + 1)(a +2) \cdots ( a+ n - 
1)$, and 
the quantities
$\mathscr{N} =1+  A^2_s/n_\mathrm{B} + A^2_s \lambda_{\mathrm{thr}}^2 
/\lambda^2$ 
and 
$\mathscr{D} = 1+  A^2_s \lambda_{\mathrm{thr}}^2/\lambda^2$.
The zero-Fock-number occupation  can be obtained from 
the normalization condition $\sum_{n =0}^{\infty} p_n = 1$, 
yielding $p_0 =\left[ {}_{2}F_{1} \left(1, \mathscr{N}, \mathscr{D}, 
\frac{n_\mathrm{B}}{n_\mathrm{B}+1} \right) \right],^{-1}$ where  
${}_{2}F_{1}(a, b; c; z)$ is the ordinary hypergeometric function. 
In the zero-temperature limit, Eq.~\eqref{eq:steadystate_distribution} becomes 
$p_n = p_0 A^{2n}_s/\mathscr{D}_n$
and the zero-Fock number occupation is 
$p_0 = {}_{1}F_1 \left(1; A^2_s\left( \lambda_{\mathrm{thr}}/\lambda \right)^2 
+1; A^2_s\right)$, 
where $_1 F_1 (a; b; z)$ is the confluent hypergeometric function. 
From Eq. \eqref{eq:steadystate_distribution} 
we can compute the average Fock number 
$\bar{n} = \sum_{n=0}^\infty n p_n$, obtaining:
\begin{equation}
	\label{eq:average_n}
	\bar{n} = A^2_s \left[1 - \left( 
	\frac{\lambda_\mathrm{thr}}{\lambda}\right)^2 \right]  +n_\mathrm{B} + 
	n_\mathrm{B}(1+n_\mathrm{B})\left(\frac{\lambda_{\mathrm{thr}}}{\lambda}\right)^2p_0.
\end{equation}
Above threshold ($\lambda \gg \lambda_\mathrm{thr}$) where we have $p_0 
\approx 0$,  and at zero temperature,
Eq. \eqref{eq:average_n} agrees with the semiclassical solution, see below 
Eq.~\eqref{eq:semiclassical_average_n}.
\subsection{Semiclassical equations in RWA}
\label{app:RWA-2}
In this Appendix we derive the set of semiclassical equations for the dynamics 
of the system in RWA.
To simplify the discussion we present the calculation in the fully polarized 
case ($P=1$) and
with $n_\mathrm{B} = 0$. 
We obtain the following set of equations: 
\begin{align}
	\label{eq:RWA_system_exact_1}
	\langle \dot{\hat{n}}_{\uparrow} \rangle &= 
	-  \Gamma_\mathrm{L}^\uparrow   (  \langle\hat{n}_\uparrow - 
	\hat{n}_{\downarrow} \rangle 
	-i \lambda \langle \hat{b} \hat{\sigma}_+ - 
	\hat{b}^\dagger  \hat{\sigma}_- \rangle + \Gamma_{\mathrm{L}}^\uparrow, 
	\nonumber \\
	%
	%
	\langle \dot{\hat{n}}_{\downarrow} \rangle &=  
	-  \Gamma_\mathrm{R}^\downarrow  \langle\hat{n}_{\downarrow} \rangle 
	+ i \lambda \langle \hat{b}  \hat{\sigma}_+	 -  
	\hat{b}^\dagger   \hat{\sigma}_- \rangle, \nonumber \\
	%
	%
	\langle \dot{\hat{\sigma}}_{+} \rangle &= \left( i \Delta \varepsilon - 
	\frac{\Gamma_\mathrm{R}^\downarrow }{2}\right) \langle \hat{\sigma}_+ 
	\rangle
	-i \lambda \langle (\hat{b} + \hat{b}^ \dagger ) \hat{\sigma}_z	\rangle, 
	\quad 
	\mathrm{and\ c.c.}, \nonumber \\
	\langle \dot{\hat{b}} \rangle &= \left(- i \omega_0 - \frac{\kappa}{2} 
	\right) \langle \hat{b} \rangle - i \lambda \langle  \hat{\sigma}_- 
	\rangle, \quad \mathrm{and\ c.c.}
\end{align}
We perform the semiclassical approximation with the replacement $\hat{b} 
\rightarrow \alpha$, 
where $\alpha = A e^{i \phi}$ is a complex number.
$A$ and $\phi$ identify the amplitude and phase of the oscillator, 
respectively. 
This is equivalent to neglecting quantum fluctuations for the harmonic 
oscillator. 
The expectation values involving both  oscillator and dot operators are thus 
factorized.
We work in a rotating frame with the replacements
$\langle \hat{\sigma}_- \rangle \rightarrow \langle \hat{\sigma}_- \rangle 
e^{-i \Delta \varepsilon t}$ 
and $\alpha \rightarrow \tilde{\alpha} e^{-i \omega_0 t}$. 
To make a connection with the notation in Sec.~III,
we set $S_x = \langle \hat{\sigma}_+ + \hat{\sigma}_- \rangle$, $S_y = -i 
\langle \hat{\sigma}_+ - \hat{\sigma}_- \rangle$, 
$S_z = \langle \hat{n}_\uparrow - \hat{n}_\downarrow \rangle$ and $p_1 = 
\langle \hat{n}_\uparrow + \hat{n}_\downarrow \rangle$.
Furthermore, by setting
\begin{equation}
	\label{eq:condition_gamma_1}
	\Gamma_\mathrm{L} = \Gamma_\mathrm{R}/2 = \Gamma,
\end{equation}
in Eq.~\eqref{eq:tunneling_rates}, the equation
for the total dot occupation $p_1$ decouples from the rest of the system and 
thus can be disregarded. Since
this condition does not alter the physics of the system, we focus on this case 
to simplify
the calculations.
With Eqs.~\eqref{eq:condition_gamma_1}, and with the
resonant condition $\Delta\varepsilon = \omega_0$, the system 
\eqref{eq:RWA_system_exact_1} becomes
\begin{align}
	\label{eq:system_RWA_1}
	\dot{S}_x &= -\Gamma S_x - 2 \lambda A  \sin \phi  S_z, \\
	\label{eq:system_RWA_2}
	\dot{S}_y &= -\Gamma S_y - 2 \lambda A    \cos \phi S_z, \\
	\label{eq:system_RWA_3}
	\dot{S}_z &= \Gamma  - \Gamma S_z + 2 \lambda A \left( \sin \phi  S_x + 
	\cos \phi  S_y \right) \, , \\
	\label{eq:system_RWA_4}
	\dot{A} &= - \frac{\kappa}{2} A + \frac{\lambda}{2} \left( - \sin \phi S_x 
	+  \cos \phi  S_y \right) \\
	\label{eq:system_RWA_5}
	\dot{\phi} &= -\frac{\lambda}{2 A} \left(  \cos \phi  S_x - \sin \phi  
	S_y\right) \, ,
\end{align}
where we have replaced the equations for $\alpha$ and $\alpha^*$ with the 
corresponding equations for $A$ and $\phi$.
The system has a steady solution (in the rotating frame) which can be  found 
by setting the time derivatives to zero.  The solution is also independent of 
the phase $\phi$ of the oscillator, which can be set to zero.
More generally, at finite polarization ($P<1$),
we obtain the nonlinear equation for the amplitude:
\begin{equation}
	\dot{A} =  - \frac{A}{2} \left[ 
	\kappa -  \frac{\frac{2 \lambda^2 P}{\Gamma} }{1 + \left( \frac{2 \lambda 
	A}{\Gamma}\right)^2}
	\right] =
	-\frac{A}{2} \left[ 
	\kappa + \gamma_\mathrm{rw}(A) 
	\right] 
	\, .
\end{equation}
In the latter equality, we have defined the effective, negative nonlinear 
damping.
When $\dot{A}=0$,
this equation yields the steady-state solutions for 
the occupation number $(\bar{n} = A^2)$ of the oscillator:
\begin{equation}
	\label{eq:semiclassical_average_n}
	\bar{n} = 0 \quad \mathrm{and} \quad 
	\bar{n} = A^2_s \left[1 - \left(\frac{\lambda_\mathrm{thr}}{\lambda} 
	\right)^2\right].
\end{equation}
with
$A^2_s = \Gamma P/(2 \kappa), \quad \lambda_\mathrm{thr} = \sqrt{\Gamma 
\kappa/(2 P)},$
and in full agreement with Eq.~\eqref{eq:nsat_threshold_p}.
The solution with $\bar{n} \neq 0$ is stable and exists only for $\lambda > 
\lambda_\mathrm{thr}$,
and corresponds to the lasing solution: for high quality factor, the 
saturation number is much larger than 1. 
The solution $\bar{n} = 0$ is stable below the threshold and unstable above it.
When $\lambda \gg \lambda_\mathrm{thr}$, $\gamma_\mathrm{rw}$ becomes 
independent of $\lambda$, 
saturating the average occupation $\bar{n}$ as a function of $\lambda$ to the 
value $A_s^2$.
For  $P=1$ but arbitrary $\Gamma_{\mathrm{L}}^\uparrow$ and 
$\Gamma_\mathrm{R}^\downarrow$, one has to include also the equation for 
$p_1$. By repeating the treatment, we obtain the expressions for 
$\gamma_{\mathrm{rw}}$, $A_s$ and $\lambda_{\mathrm{thr}}$ given in Sec.~III.
%
%
%
%
%
%
\section{Semiclassical equations beyond RWA}
\label{app:beyondRWA}
We derive here the set of semiclassical equations for the dynamics of the 
system, starting from 
the full Hamiltonian Eq.~\eqref{eq:system_hamiltonian}.
Using Eq.~\eqref{eq:masterequation}, we obtain the following set of exact 
equations 
\begin{equation}
	\label{eq:RWA_system_exact_2}
	\begin{split}
		\langle \dot{\hat{n}}_{\uparrow} \rangle &= - 
		\Gamma_\mathrm{L}^\uparrow  	\langle\hat{n}_{\uparrow} \rangle 
		- \Gamma_\mathrm{L}^\uparrow\langle\hat{n}_{\downarrow} \rangle   -i 
		\lambda \langle (\hat{b} + \hat{b}^\dagger )(\hat{\sigma}_+ - 
		\hat{\sigma}_-) \rangle  + \Gamma^\uparrow_{\mathrm{L}}, \\
		%
		%
		\langle \dot{\hat{n}}_{\downarrow} \rangle &=     - 
		\Gamma_\mathrm{R}^\downarrow  \langle\hat{n}_{\downarrow} \rangle +i 
		\lambda \langle (\hat{b} 
		+ \hat{b}^\dagger ) (\hat{\sigma}_+	 -  \hat{\sigma}_-) \rangle 
		, \\
		%
		%
		\langle \dot{\hat{\sigma}}_{-} \rangle &= \left(- i \Delta \varepsilon 
		- \frac{\Gamma_\mathrm{R}^\downarrow}{2}\right) \langle \hat{\sigma}_- 
		\rangle
		+i \lambda \langle (\hat{b} + \hat{b}^ \dagger ) \hat{\sigma}_z	\rangle
		, \quad \mathrm{and\ c.c.},  \\
		%
		%
		\langle \dot{\hat{b}} \rangle &= \left(- i \omega_0 - \frac{\kappa}{2} 
		\right) \langle \hat{b} \rangle - i \lambda \langle (\hat{\sigma}_+ + 
		\hat{\sigma}_- \rangle
		\,   \quad \mathrm{and\ c.c.}. 
	\end{split}
\end{equation}
We perform again the semiclassical approximation and move to the rotating 
frame; assuming the  condition Eq.~(\ref{eq:condition_gamma_1}), 
the equation for the total dot occupation $p_1$ still decouples from the rest 
of the system. 
\subsection{Resonant case}
On resonance  ($\Delta\varepsilon = \omega_0$) and for $P=1$ the system 
\eqref{eq:RWA_system_exact_2} becomes
\begin{alignat}{3}
	\label{eq:system_1}
	\dot{S}_x &= &&-\Gamma S_x - 2 \lambda A \left[\sin (2\omega_0 t - \phi) + 
	\sin \phi \right] S_z, \\
	\label{eq:system_2}
	\dot{S}_y &= &&-\Gamma S_y - 2 \lambda A \left[\cos(2\omega_0 t - \phi) + 
	\cos \phi \right] S_z, \\
	\label{eq:system_3}
	\dot{S}_z &= &&- \Gamma S_z + 2 \lambda A \left\{ \left[\sin(2\omega_0 t - 
	\phi) + \sin \phi \right] S_x \right. \nonumber \\
	& &&+\left. \left[\cos(2\omega_0 t - \phi) + \cos \phi \right]S_y \right\} 
	+ \Gamma, \\
	\label{eq:system_4}
	\dot{A} &= &&- \frac{\kappa}{2} A + \frac{\lambda}{2} \left\{\left[ \sin(2 
	\omega_0 t - \phi ) - \sin \phi\right]S_x \right. \nonumber \\
	& &&+\left. \left[\cos(2\omega_0 t - \phi )+ \cos \phi \right]S_y 
	\right\}, \\
	\label{eq:system_5}
	\dot{\phi} &= &&-\frac{\lambda}{2 A} \left\{ \left[ \cos (2\omega_0 t - 
	\phi) + \cos \phi \right] S_x \right. \nonumber \\
	& &&- \left.\left[ \sin( 2\omega_0 t - \phi) + \sin \phi \right] 
	S_y\right\}.
\end{alignat}
We have now  terms rotating at frequency $2\omega_0$ in the system. 
It is possible to obtain a single recursive equation for the
Fourier coefficients of $S_z$, which is related to the nonlinear damping 
$\gamma_\mathrm{eff}$, as follows: 
we first assume that the amplitude $A$ of the oscillator in 
Eqs.~\eqref{eq:system_1}-\eqref{eq:system_3} for 
the spin dynamics is constant. This assumption is based on 
the separation of timescales  $\kappa \ll \Gamma, \lambda, \omega_0$, 
which guarantees that the amplitude of the oscillations is indeed a slow 
variable coupling only to the 
average spin over the time evolution 
in the rotating frame. 
Furthermore, we can disregard the evolution of the phase $\phi$ as for the RWA 
case. 
With these assumptions, we can focus on 
Eqs.~\eqref{eq:system_1}-\eqref{eq:system_3} for the spin degrees
of freedom alone. They can be cast in the form reported in 
Eqs.~\eqref{eq:system_beyond_RWA}-\eqref{eq:effective_field}.
We consider the Fourier expansion in 
harmonics of the fundamental frequency $2\omega_0$ of the spin quantities, 
i.e.:
\begin{equation}
	\label{eq:fourier_coefficients}
	S_k (t) = \sum_{n = -\infty}^{\infty} S_k ^{(n)} (A) e^{2 i n \omega_0 
	t},\qquad (k = x, y, z),
\end{equation}
where we have made explicit the amplitude dependence of the Fourier 
coefficients.
By plugging Eq.~\eqref{eq:fourier_coefficients} in 
Eqs.~\eqref{eq:system_1}-\eqref{eq:system_3} 
we are able to write a single
equation for $S_z^{(n)}$, which couples to $S_z^{(n+1)}$ and $S_z^{(n-1)}$. It 
reads:
%
%
%
\begin{align}
	\label{eq:recursive_equation}
	&\left[\chi^{-1}_n + \left(\frac{2 \lambda A}{\Gamma}\right)^2 
	\left(\chi_{n} 
	+ \frac{\chi_{n-1} + \chi_{n+1}}{2}\right) \right] S_z^{(n)} =  \nonumber 
	\\
	&=\delta_{n,0} 
	-\left(\frac{2 \lambda A}{\Gamma}\right)^2 \left(\frac{\chi_{n} + 
		\chi_{n+1}}{2} S_z^{(n+1)} + \frac{\chi_{n-1} + \chi_{n}}{2} 
	S_z^{(n-1)}\right),
\end{align}
where we introduced the generalized dimensionless susceptibility $\chi_n = 
\Gamma/(\Gamma + 2 i n \omega_0)$.
Equation \eqref{eq:recursive_equation} constitutes a matrix equation with an 
infinite band-diagonal matrix,
having only three non-zero diagonals, and a constant vector. It can be solved 
numerically by
truncating the resulting matrix since the Fourier coefficients decay rapidly 
for increasing $n$. 
After solution of Eq.~\eqref{eq:recursive_equation}, we can find $S_x^{(n)}$ 
and $S_y^{(n)}$ in terms of $S_z^{(n)}$, plug them into 
Eq.~\eqref{eq:system_4} and derive the nonlinear damping as given by 
Eq.~\eqref{eq:gammaeff}.
%
%
For $\lambda A, \Gamma \ll \omega_0$, Eq.~\eqref{eq:gammaeff} agrees with the 
result of the RWA, where all harmonics with $n>0$ vanish
and the system has a steady solution in the rotating frame.
As the effective Rabi frequency $\lambda A$ increases, energy is fed into
higher harmonics of $S_z$, as a result of the nonlinear interaction between
the oscillator and the spin degrees of freedom. This produces a nonmonotonic
behavior in $\gamma_\mathrm{eff}$ as a function of $\lambda A$, which 
is responsible for the appearance of multiple stable limit cycles in the 
oscillator amplitude.
\subsection{Off-resonant case}
\label{app:beyondRWA-2}
The treatment can be readily generalized to the off-resonant case, where
$\Delta\varepsilon \neq \omega_0$.
In this case  
the recursive equation satisfied by the Fourier coefficients of $S_z$ reads 
\begin{align}
	\label{eq:recursive_equation_nonresonant}
	&\left[\!\frac{(\chi_{n-1}^-)^{-1} \!+\!(\chi_{n}^+)^{-1}}{2}\!\! +\!\! 
	\left(\frac{2 \lambda A}{\Gamma}\right)^2\!\! \left(\frac{\chi_{n}^+ \!+\! 
		\chi_{n}^- \!+\!\chi_{n-1}^+ \!+\! \chi_{n-1}^- }{2} 
		\right)\right]\!\! 
	S_z^{(n)} \!\!=\nonumber \\  &=\delta_{n,0} \!\!-\!\! \left(\frac{2 
	\lambda 
		A}{\Gamma}\right)^2 \left(\frac{\chi_{n}^+ \!+\! 
		\chi_{n}^-}{2}S_z^{(n+1)} + 
	\frac{\chi_{n-1}^+\! +\! \chi_{n-1}^-}{2}S_z^{(n-1)} \right),
\end{align}
with the generalized susceptibilites
\begin{align}
	\chi_n^- &= \frac{\Gamma}{\Gamma + i[2\omega_0 n + (\omega_0 - 
	\Delta\varepsilon)]},\nonumber \\
	\chi_n^+ &= \frac{\Gamma}{\Gamma + i[2\omega_0 n + (\omega_0 + 
	\Delta\varepsilon)]}.
\end{align}
For $\Delta\varepsilon \rightarrow \omega_0$, we have $\chi_n^- \rightarrow 
\chi_n$
and $\chi_n^+ \rightarrow \chi_{n+1}$, and we recover 
Eq.~\eqref{eq:recursive_equation}.
The nonlinear damping for the amplitude is then given by 
\begin{align}
	\label{eq:gamma_eff_nonresonant}
	\gamma_\mathrm{eff}(A) = -\frac{ 2\lambda^2}{\Gamma} &\left \{\frac{4 
	\omega_0 \Delta\varepsilon}{\Gamma^2 \left( 1 + \frac{\Delta\varepsilon^2 
	- \omega_0^2}{\Gamma^2}\right)^2 + 4\omega_0^2} S_z^{(0)} \right. 
	\nonumber \\ &\left.- \mathrm{Im} \left[ \frac{2 \Delta\varepsilon}{\Gamma 
	\left(1 + \frac{\Delta\varepsilon^2 - \omega_0^2}{\Gamma^2} \right) + 2 i 
	\omega_0} S_z^{(1)}\right] \right\}.
\end{align}
The expression is in agreement with Eq.~\eqref{eq:gammaeff} 
when $\Delta\varepsilon = \omega_0$.
We have used Eq.~\eqref{eq:recursive_equation_nonresonant} together
with Eq.~\eqref{eq:gamma_eff_nonresonant} to generate
the semiclassical stability diagram of Fig.~\ref{fig:3}(a).
%
%
%
%
%
%
\section{Current and shot-noise using the full counting statistics method}
\label{app:FCS}
We report here the procedure for the numerical calculation for the average 
current and the zero-frequency current noise (shot-noise) through the quantum 
dot. 
We employ the full counting statistics (FCS) method (see, for instance, Refs. 
\onlinecite{PhysRevB.70.205334,Flindt2005,PhysRevB.82.155407}).
To express the average current 
$I$
and the zero-frequency noise 
$S(0)$, we use a vector a representation
for the Hilbert-space operators: the Liouvillian superoperator $\mathcal{L}$ 
operates in the Liouville space, 
where a Hilbert-space operator $\hat{A}$ is represented by a vector
$|a \rrangle$, and premultiplication (left) or postmultiplication (right)
of $\hat{A}$ are represented by an appropriate matrix which multiplies 
the vector $|a \rrangle$. 
The Liouville space possesses a natural
scalar product given by $\llangle a | b \rrangle = \mathrm{Tr} 
(\hat{A}^\dagger \hat{B} )$,
where the trace is performed over the Hilbert space. In this way,
the master equation Eq.~\eqref{eq:WBR} reads   $|\dot{\rho} \rrangle = 
\mathcal{L} | \rho \rrangle$. Since the Liouvillian is in general 
non-Hermitian, it has different left and right eigenvectors, namely
\begin{equation}
	\mathcal{L} |r_i \rrangle = \lambda_i |r_i \rrangle, \quad \llangle l_i | 
	\mathcal{L} = \lambda_i \llangle l_i |.
\end{equation}
We denote with $| \rho_\mathrm{st} \rrangle$ the steady-state of the system, 
which satisfies the
equation $\mathcal{L}| \rho_\mathrm{st} \rrangle = 0$ and hence constitutes 
the right eigenvector corresponding
to the eigenvalue $\lambda_0 = 0$ of the Liouvillian. The left eigenvector is 
readily found from the 
orthonormality condition $\mathrm{Tr} (\hat{\rho}_\mathrm{st}) = 1 = 
\mathrm{Tr} (\hat{\mathds{1}}^\dagger\hat{\rho}_\mathrm{st}).$ Hence, the left 
eigenvector corresponds
to the identity operator in Hilbert space, which we denote with $\llangle 
\mathds{1} |$.
Next, in the framework of the FCS, we define the collector in our system to be 
the right lead 
(in the large bias limit only left-to-right transport is allowed). 
The  current superoperator is then defined by
\begin{equation}
	\mathcal{J}| \rho \rrangle  = 
	\sum_{\sigma} \Gamma^\sigma_\mathrm{R} \hat{F}_\sigma \hat{\rho} 
	\hat{F}^\dagger_\sigma.
\end{equation}
With this definition, the average current reads
\begin{equation}
	\label{eq:average_current}
	I = e \llangle \mathds{1} | \mathcal{J} | \rho_\mathrm{st} \rrangle = 
	\mathrm{Tr} ( \mathcal{J}\hat{\rho}_\mathrm{st} ) = e \sum_{\sigma} 
	\Gamma^\sigma_\mathrm{R} \rho^\mathrm{st}_{\sigma},
\end{equation}
where $\rho^\mathrm{st}_{\sigma}$ 
is the occupation probability of the spin-$\sigma$ level in the dot, in the 
steady-state. Equation~\ref{eq:average_current} corresponds to 
Eq.~\eqref{eq:current}. For the zero-frequency current 
noise, one finds \cite{Flindt2005}
\begin{align}
	\label{eq:current_noise}
	S(0) &=e^2 \llangle \mathds{1} | \mathcal{J} | \rho_\mathrm{st} \rrangle 
	-2 e^2  \llangle \mathds{1} | \mathcal{J} \mathcal{R}\mathcal{J}  | 
	\rho_\mathrm{st} \rrangle \nonumber \\
	&=e I- 2e^2  \Tr (\mathcal{J} \mathcal{R}\mathcal{J} 
	\hat{\rho}_\mathrm{st}).
\end{align}
We have introduced the pseudoinverse of the Liouvillian $\mathcal{R} = 
\mathcal{Q} \mathcal{L}^{-1} \mathcal{Q}$, 
where $\mathcal{Q}$ is the projector out of the null-space of $\mathcal{L}$, 
which is spanned by $\hat{\rho}_\mathrm{st}$. 
If $| \rho_\mathrm{st} \rrangle \llangle \mathds{1}|$ is the projector onto 
the stationary state then $\mathcal{Q} = \mathds{1} - | \rho_\mathrm{st} 
\rrangle \llangle \mathds{1}|$. 
The pseudoinverse $\mathcal{R}$ is well defined, since the inversion is 
performed in 
the subspace spanned by $\mathcal{Q}$, where $\mathcal{L}$ is regular.
%
%
%
%
%
%
%
\section{Current for the two-state model in the bistability regime}
\label{app:TSM}
%
%
%
\begin{figure}[b]
	\includegraphics[scale=0.55]{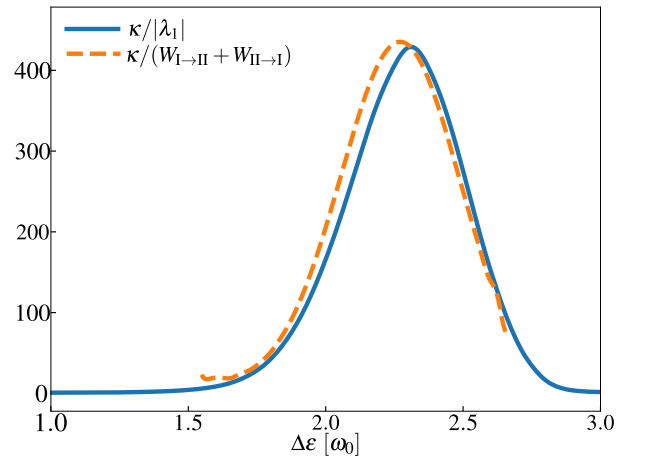}
	\caption{ Comparison between the inverse of the smallest nonzero 
		eigenvalue  $|\lambda_1|$ of the Liouvillain (solid blue line) and 
		$(W_{\mathrm{I}\rightarrow\mathrm{II}} + 
		W_{\mathrm{II}\rightarrow\mathrm{I}})^{-1}$ as  obtained by the 
		two-state model 
		(orange dashed line), as a function of the dot's energy splitting 
		$\Delta\varepsilon$. 
		The curves are rescaled with the typical decay time of the oscillator 
		$\kappa^{-1}$.  
		Parameters: $Q = 10^3,\ \Gamma_{\mathrm{L}}= \Gamma_{\mathrm{R}} = 
		0.1\omega_0$, $P=1$, $\lambda = 0.13\omega_0$, $T=0$.}
	\label{fig:5}
\end{figure}
%
%
%
The two-state approximation for the system is valid if the $p_n$ distribution 
of the oscillator
displays two distinct peaks of similar probability $\mathcal{P}_\mathrm{I}$ and
$\mathcal{P}_\mathrm{II}$, which are well separated by a region with a 
negligible
probability, as is reported in Fig.~\ref{fig:1}(e). To show
telegraph noise, it is also necessary that the 
current variance associated to each state is smaller than the distance between 
the average values, i.e., $\Delta I_\mathrm{I},\ \Delta I_\mathrm{II}  < 
|I_\mathrm{I}- I_\mathrm{II}|$, and that the switching rates 
between the two states are slow, such that one can resolve the individual 
jumps by monitoring
the current during time.
Under these conditions, we can model the current and the current noise by 
using a set of four parameters,
$I_\mathrm{I}$, $I_\mathrm{II}$ and the rates 
$W_{\mathrm{I}\rightarrow\mathrm{II}}$ and $W_{\mathrm{II}\rightarrow 
\mathrm{I}}$. The two states  will have relative probabilities
\begin{equation}
	\mathcal{P}_\mathrm{I} = 
	\frac{W_{\mathrm{II}\rightarrow\mathrm{I}}}{W_{\mathrm{I}\rightarrow\mathrm{II}}+W_{\mathrm{II}\rightarrow
	 \mathrm{I}}},\quad \mathcal{P}_\mathrm{II} = 
	\frac{W_{\mathrm{I}\rightarrow\mathrm{II}}}{W_{\mathrm{I}\rightarrow\mathrm{II}}+W_{\mathrm{II}\rightarrow
	 \mathrm{I}}}.
\end{equation}
The average current and the zero-frequency current noise are given by 
\begin{equation}
	\label{eq:average_current_ts}
	I_{ts}= \frac{W_{\mathrm{II}\rightarrow\mathrm{I}} I_\mathrm{I} 
	+W_{\mathrm{I}\rightarrow \mathrm{II}} I_\mathrm{II} 
	}{W_{\mathrm{I}\rightarrow\mathrm{II}}+W_{\mathrm{II}\rightarrow 
	\mathrm{I}}}
\end{equation}
and 
\begin{equation}
	\label{eq:current_noise_ts}
	S(0)_{ts} = \frac{4 \mathcal{P}_\mathrm{I} \mathcal{P}_\mathrm{II} 
	(I_\mathrm{I} - 
	I_\mathrm{II})^2}{W_{\mathrm{I}\rightarrow\mathrm{II}}+W_{\mathrm{II}\rightarrow
	 \mathrm{I}}},
\end{equation}
where the numerator is 
the two-state current variance [\onlinecite{PhysRevB.78.024513}].
To calculate these quantities in our system, we identify 
$\mathcal{P}_\mathrm{I}$ and  $\mathcal{P}_\mathrm{II}$ 
with the area of each of the two peaks in the steady-state $p_n$ distribution 
of the oscillator; next we set the elements
of the density matrix corresponding to one of the two states to zero, and we 
build a new truncated density matrix from which
one can calculate the two currents $I_\mathrm{I}$ and $I_\mathrm{II}$ through 
Eq.~\eqref{eq:average_current}, hence
the average current with Eq.~\eqref{eq:average_current_ts}. The 
current variance for each state can be estimated as $\Delta I_\mathrm{I, II} = 
\sqrt{\Gamma S_\mathrm{I,II}}$, where $S_{\mathrm{I,II}}$ is the 
zero-frequency noise 
calculated from Eq.~\eqref{eq:current_noise}, but using the truncated states. 
The sum of the rates 
$W_{\mathrm{I}\rightarrow\mathrm{II}}+W_{\mathrm{II}\rightarrow \mathrm{I}}$ 
is obtained by comparing the current noise
calculated with Eq.~\eqref{eq:current_noise} with the one given by 
Eq.~\eqref{eq:current_noise_ts}.
In the two-state model, a very slow timescale dominates the current noise. 
Specifically, this slow timescale is associated with the real part of the 
smallest nonzero eigenvalue of the Liouvillian of the system, as one can see 
directly by expanding Eq.~\eqref{eq:current_noise} in terms of the eigenvalues 
and eigenvectors of $\mathcal{L}$ [\onlinecite{PhysRevB.78.024513}]. If the 
lowest nonzero eigenvalue, $\lambda_1$, is small and well separated from the 
others (i.e., $|\lambda_1| \ll |\lambda_p|$ for $p>1$), the current noise is 
dominated by this eigenvalue, and a comparison
with Eq.~\eqref{eq:current_noise_ts} leads us to identify $-\lambda_1 = 
W_{\mathrm{I}\rightarrow\mathrm{II}}+W_{\mathrm{II}\rightarrow \mathrm{I}}$. 
In Fig.~\ref{fig:5} 
we compare the result for the sum of the rates obtained by the eigenvalue 
expansion with the two-state approximation,
showing that the behavior is very similar. Moreover, this timescale is much 
larger compared to the relaxation time of the oscillator, and shows indeed 
that the 
telegraph  dynamics can be observed by monitoring the current.
We stress that here we show relatively large coupling constants in order to 
realize numerical calculations with Fock occupation number not too large.
On the other hand, semiclassical equations at finite polarization predict a 
similar behavior 
also at smaller coupling constants. 
%
%
%
%
%
\begin{figure}[t!]
	\includegraphics[width=\columnwidth]{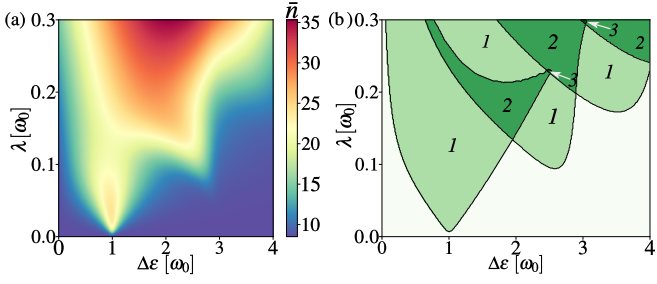}
	\caption{Effect of finite temperature and finite polarization.
		(a) Average occupation number in the oscillator at resonance 
		as a function of the dot's energy splitting $\Delta\varepsilon$ and of
		the spin-oscillator coupling strength $\lambda$. (b) Stability 
		diagram of the oscillator. The italic numbers indicate
		the number of distinct peaks in the Fock distribution. Parameters: $Q 
		= 10^3,\ \Gamma_{\mathrm{L}}= \Gamma_{\mathrm{R}} = 0.1\omega_0$, 
		$P=0.5$, $T=10\omega_0$.}
	\label{fig:6}
\end{figure}
%
%
%
%
%
%
%
%
\section{Multistability in nonideal cases}
\label{app:nonideal}
\subsection{Effect of finite temperature $(T>0)$ and finite polarization 
$(P<1)$}
\label{app:nonideal1}
The model system we considered can be implemented in a nanomechanical 
framework, by
considering for example a carbon nanotube quantum dot (CNTQD). 
Mechanical resonators have in general low frequency ($\omega_0/2\pi \approx 
\SI{100}{MHz}$),
and consequently one cannot neglect the effect of finite temperature of the 
thermal bath
coupled to them, since $T \gtrsim \omega_0$. Furthermore, state-of-the-art 
ferromagnetic contacts reach a polarization of about 40-50\%, thereby 
decreasing the 
lasing efficiency. In Fig.~\ref{fig:6} we report the numerical calculation of 
the average occupation of the oscillator in the steady-state---obtained with 
Eq.~\eqref{eq:master_equation}---together with the 
stability diagram for a nonideal case ($T \gg \omega_0$ and $P<1$), and we 
show how the qualitative picture is not destroyed. More specifically, the 
lasing threshold is pushed to a larger coupling, according to 
Eq.~\eqref{eq:nsat_threshold_p}, as well as the onset of bi- and 
multistability. The thermal noise smears out the transitions to the lasing 
state.
\subsection{Effect of spin relaxation at $T>0$ and $P<1$}
\label{app:nonideal2}
We take into account decoherence in the quantum dot 
due to spin relaxation with a characteristic time $T_1$.
We neglect a general inhomogeneous pure dephasing term of characteristic
timescale $T_\phi$, which is justified as this term arises from hyperfine 
coupling of the electronic spin to the nuclear spin of $^{13}\mathrm{C}$ 
atoms, 
whose natural abundance in carbon is less than 1\% 
[\onlinecite{Churchill:2009fy}]. 
The spin relaxation is included in the dynamics by adding the dissipator 
$\mathcal{L}_\mathrm{sr} \hat{\rho} = \gamma_\mathrm{sr} 
\mathcal{D}(\hat{\sigma}_-) \hat{\rho}$
%
%
to Eq.~\eqref{eq:master_equation}.
$\gamma_\mathrm{sr} = T_1^{-1}$ identifies the relaxation rate. 
Spin relaxation plays a role similar to the effect of finite polarization: an 
electron 
decays into the lower spin level and then tunnels into the right lead, without 
emitting 
a quantum of oscillation. If the relaxation rate is much smaller than the Rabi 
frequency $\lambda A$ and of the tunneling rates, the dynamics is expected to 
be unperturbed.
We find numerically the steady-state for the new Lindblad equation, 
and we calculate the average Fock number of the oscillator for different 
values 
of $\gamma_\mathrm{sr}$.
An example is  shown in Fig.~\ref{fig:7}(a) in which 
the lasing mechanism is noticeably suppressed only for $ \gamma_\mathrm{sr} 
/\omega_0 = \num{e-2}$.
Figures \ref{fig:7}(b) and \ref{fig:7}(c) show the average occupation and the stability 
diagram as a function of $\lambda$ and $\Delta\varepsilon$.
For the case of a CNTQD setup, the  relaxation time in 
single-walled CNTs [\onlinecite{Rice:2013ep}]  was reported to be $T_1 \approx 
\SI{100}{\micro\second}$ at $T = \SI{4}{\kelvin}$ 
corresponding to a relaxation rate of $10 \,\, \mbox{kHz}$.
At low temperature ($T \approx 20\,\, \mbox{mK}$, considered in our case)
we expect a substantial decrease of this value. 
\subsection{Effect of nonlinearity at $T>0$ and $P<1$}
\label{app:nonideal3}
We include in our numerical model a Duffing nonlinearity for the harmonic 
oscillator, 
by modifying Hamiltonian~\eqref{eq:system_hamiltonian} into
\begin{equation}
	\hat{H} = \frac{\Delta \varepsilon}{2}\hat{\sigma}_z
	+ \omega_0 \hat{b}^\dagger \hat{b} + \frac{\tilde{\beta}}{4}(\hat{b} + 
	\hat{b}^\dagger)^4
	+ \lambda (\hat{\sigma}_+ + \hat{\sigma}_-)  (\hat{b} + \hat{b}^\dagger).
\end{equation}
We introduced the parameter $\tilde{\beta} = \beta x^4_\mathrm{ZPM}$, 
with $\beta$ and $x_\mathrm{ZPM} = \sqrt{\hbar/2 m \omega_0}$ being the 
Duffing 
nonlinearity parameter and the zero-point amplitude of the oscillator (where 
we have restored $\hbar$), respectively. 
The nonlinearity is expected to play a nonneglibile role for mechanical 
resonators where intrinsic nonlinearities can be large and hence might affect 
the lasing behavior at large amplitudes.
For a realistic estimate of $\tilde{\beta}$, we set
the typical mass of a CNT to be $m \approx \num{e-21}$ kg, which for 
$\omega_0/2\pi = \SI{100}{MHz}$ 
gives zero-point fluctuations of order $x_\mathrm{ZPM} \approx 
\SI{10}{\pico\meter}$.
Experimentally,
the geometrical nonlinearity 
parameter for a CNT is positive and of order 
$\beta/m = \SI{e35}{\newton\per\kilogram\per\cubic\meter}$  
[\onlinecite{Meerwaldt2012}].
The parameter $\tilde{\beta}/2\pi$ is hence of order $\approx \SI{1}{kHz}$, 
i.e., $\tilde{\beta}/\omega_0 \approx \num{e-5}$.
We neglect the electrostatic nonlinearity arising from strong coupling effects 
between the leads and the CNT and from single-electron tunneling, 
which is in general orders of magnitude smaller and is proportional to the 
electron tunneling rate, assumed much smaller than $\omega_0$.
Solving the Lindblad equation for the steady-state, we  report  the average 
Fock number as a function of the coupling strength $\lambda$ 
in Fig.~\ref{fig:8}(a). 
Finally, in Fig.~\ref{fig:8}(b,c) we show the average occupation and the 
stability diagram as a function of $\lambda$ and $\Delta\varepsilon$ 
by combining the effect of finite temperature, finite polarization, spin 
relaxation and Duffing nonlinearity showing that 
the main features still persist in a largely nonideal case.

\begin{widetext}
	%
	%
	\begin{figure*}[t]
		\includegraphics[width=0.9\textwidth]{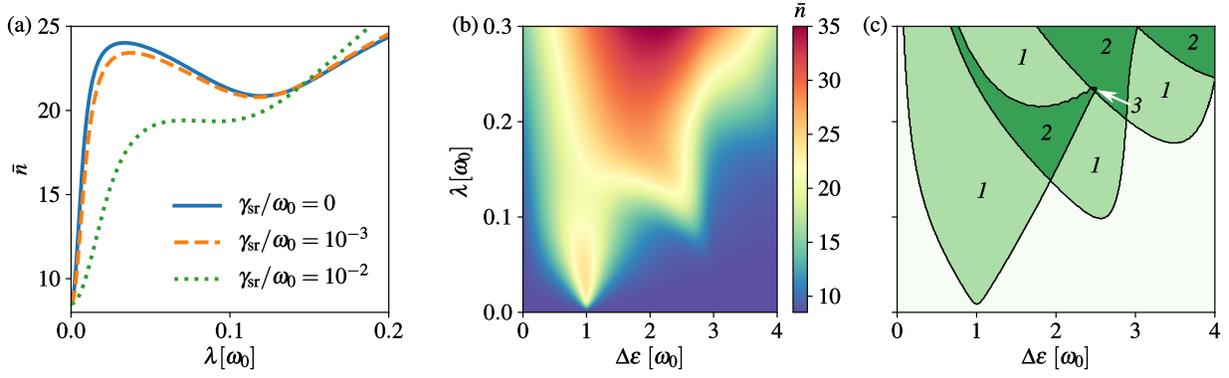}
		\caption{
			Effect of dot's spin relaxation.
			(a) Average occupation of the oscillator on resonance  as a 
			function of 
			$\lambda$ at different values of the spin relaxation rate 
			$\gamma_{\mathrm{sr}}$.
			(b) Average occupation of the oscillator as a function of 
			$\lambda$  and 
			of $\Delta\varepsilon$ for $\gamma_{\mathrm{sr}}=10^{-3}\omega_0$.
			(c) Stability diagram of the oscillator for 
			$\gamma_{\mathrm{sr}}=10^{-3}\omega_0$: the italic numbers 
			indicate the 
			number of distinct peaks in the Fock distribution.
			Parameters: $Q=10^{3}$, 
			$\Gamma_\mathrm{L}=\Gamma_\mathrm{R}=0.1\omega_0$, 
			$P=0.5$, $T= 10\omega_0$.
		}
		\label{fig:7}
	\end{figure*}
	%
	%
	%
	%
	\begin{figure*}[t]
		\includegraphics[width=0.9\textwidth]{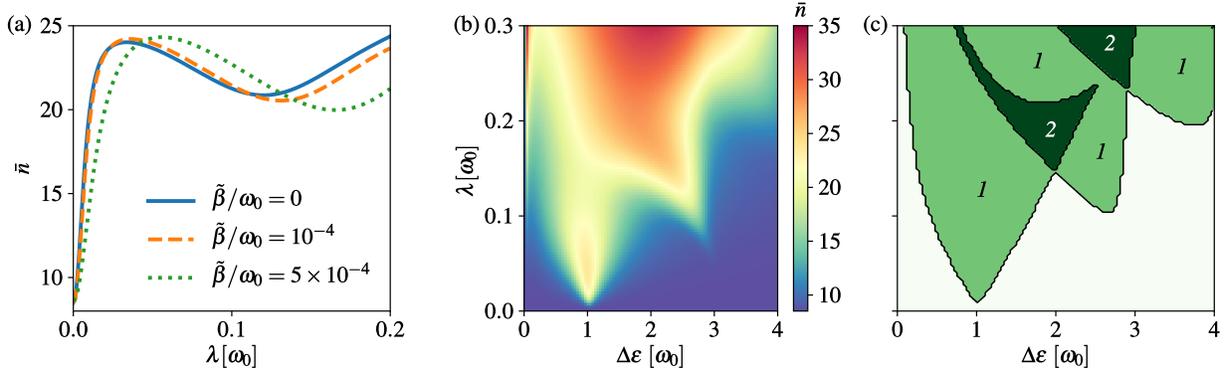}
		\caption{
			Effect of the Duffing nonlinearity for a nanomechanical resonator. 
			(a) Average occupation for the oscillator as a function of 
			$\lambda$ for three different values of the Duffing nonlinearity 
			parameter and $\gamma_{\mathrm{sr}}=0$.
			(b) Average occupation for the oscillator as a function $\lambda$ 
			and $\Delta\varepsilon$ for $\gamma_{\mathrm{sr}}=10^{-3}\omega_0$ 
			and $\tilde{\beta}= 10^{-4}\omega_0$.
			(c) Stability diagram of the oscillator for 
			$\gamma_{\mathrm{sr}}=10^{-3}\omega_0$ and $\tilde{\beta}= 
			10^{-4}\omega_0$: the italic numbers indicate the number of 
			distinct peaks in the Fock distribution.
			Parameters: $Q=10^{3}$, 
			$\Gamma_\mathrm{L}=\Gamma_\mathrm{R}=0.1\omega_0$, $P=0.5$, $T= 
			10\omega_0$.
		}
		\label{fig:8}
	\end{figure*}
	%
	%
	%
\end{widetext}
%

%
%
%
%
%
%
%
%

%
\end{document}